\documentclass[dvips,12pt]{article}
\usepackage{graphicx}
\catcode`\@=11
\begin{document} \openup6pt
%%%%%%%%%%%%%%%%%%%%%%%%%%%%%%
\title{ VISCOUS COSMOLOGIES  WITH  VARIABLE  $ G$  AND $\Lambda $ IN $R^2$ GRAVITY }
\author{B. C. Paul$^1$\thanks{E-mail: bcpaul@iucaa.ernet.in } and  P. S. Debnath$^2$\thanks{E-mail: parthasarathi6@hotmail.com}\\
$^1$Physics Department, North Bengal University \\
  Siliguri, Pin : 734 013, India. \\
 $^2$Physics Department, A. B. N. Seal College,  \\
  Coochbehar, Pin : 736 101, India.  }
  \date{}
\maketitle

\begin{abstract} We study  evolution of a flat Friedmann-Robertson Walker universe
filled with a bulk viscous cosmological fluid in a higher derivative theory of gravity
  in the presence of time varying gravitational  and cosmological constant.
  Cosmological models admitting both power-law and 
exponential expansions  are   explored here in the presence of imperfect
fluid  described  by 
 full Israel and Stewart   theory. We note some new and interesting
 cosmological solutions relevant for model building including present accelerating phase. In the case of power law,
  it is found that gravitational constant increases as the time
evolves for a positive cosmological constant whereas it decreases for a negative
cosmological constant.  The evolution of temperature of a viscous universe
is  also determined. \\
PACS No(S): 98.80. Cq
\end{abstract}

\vspace{1.5in}
\pagebreak

%%%%%%%%%%%%%%%%%%%%%%%%
\section{Introduction}
%%%%%%%%%%%%%%%%%%%%%%%%
Recent astronomical observations of type Ia supernovae 
 with redshift parameter $Z \leq 1$ \cite{f1}, Wilkinson Microwave
 Anisotropy Probe (WMAP) \cite{f2} etc. provided evidence that we may live
 in a low mass density universe  ($\Omega \sim 0.3$) \cite{f3}. The
 predictions of the observations lead to a convincing belief in
 modern cosmology that a part of the  universe is  filled up with dark
 energy ($\Omega \sim 0.7$), which  may
 be addressed by a suitable cosmological constant. One of the recent predictions  
 that the present universe  accelerating is  remarkable which is not  properly understood yet. The  present accelerating phase of the universe 
 may be accommodated in cosmological models, obtained from theories :  either by a modification of  the Einstein's 
theory of gravity (GR)  or  by including  a time varying cosmological
constant.  A number of literature appeared which addressed the   present 
accelerating phase making use of exotic kind of fields in the  matter sector of the Einstein's field equation. The study of late universe however, remains open to address the present issues in other theories. 
  One of the early attempts
to modify GR is based on adding curvature squared terms to the
Einstein-Hilbert action  which  is known as 
generalized theory of gravity was used to obtain early inflation. The theory is found to have a number of good features important for understanding early universe.   Starobinsky \cite{f4}  shown that $R^2$ term in the Einstein Hilbert action  admits inflation long before the advent of inflation is actually realized. However, the efficacy of the theory is known only after the seminal  work on inflation 
 by Guth \cite{f5}, who employed  temperature dependent phase transition
 mechanism to obtain inflation. The  inflationary  scenario 
of the early universe is attractive and which may solve some of the outstanding problems in cosmology \cite{f6}.  It is  known that higher order gravity  with suitable counter terms viz., $ C^{\mu\nu\rho\delta}C_{\mu\nu\rho\delta}$, 
$ R^2$, and cosmological 
constant ($\Lambda$) added to the Einstein-Hilbert action, one gets a perturbation theory 
which is well behaved, formally renormalizable and asymptotically free \cite{f7,f8}. 

Cosmological models considering  perfect fluid as a source of matter in the framework 
of higher derivative gravity have been studied in the literature \cite{f9} in order to obtain viable cosmological scenario of the early 
universe. In the absence of particle creation, particle number is conserved in the perfect fluid
 ( i.e., $ n^{\alpha}_{;\alpha} =0$, where $n$ is particle number density ). It may be pointed out here that perfect fluid in equilibrium generates neither  entropy ($S^{\alpha}_{;\alpha} =0$) nor frictional heat flow as  their dynamics is reversible.

 However, in the early universe a  number of processes might have 
occurred  leading to  a deviation from perfect
fluid assumption e.g. viscosity which is to be taken into account. It is known that  real fluids behave irreversibly and  therefore it is important to consider dissipative processes both   
 in cosmology and in astrophysics. Some of the  dissipative processes in the
early universe  responsible for viscosity may be  due to   
  the decoupling of neutrinos from  the 
radiation era, the decoupling of matter from  radiation during the 
recombination era, creation of  superstrings in the quantum era, 
particle collisions involving gravitons, cosmological quantum particle 
creation processes and  formation of galaxies \cite{f10}. 
 It has been predicted from observations    that a non negligible dissipative bulk stress on
   cosmological scales at the 
late universe  phase might be important. The possible source of 
 such viscosity may be due to (i) gaseous matter in the framework of relativistic gravity which 
  may give rise to internal 
  self-interaction leading to a negative cosmic bulk pressure \cite{f11},
  (ii) deviation  of the non relativistic particle in the substratum
  from dust. For a  non-relativistic substratum cosmic anti-friction may generate a negative fluid bulk pressure which 
 has been noted \cite{f12} in the framework of Einstein gravity. 

    Since the pioneering work of Dirac \cite{f13}, who proposed a theory
    with a time varying  gravitational coupling  constant $G(t)$, a
    number of literature \cite{f14} appeared with variable $G(t)$ and
    $\Lambda(t)$ in higher derivative theory of gravity in order  to  obtain
    cosmological models accommodating  the present cosmic acceleration.  Recent applications of the 
apparent magnitude-redshift test, based on type Ia supernovae \cite{f2,f15}, strongly 
favours a theory with  a positive $\Lambda $ term.  A number of ansatze have been proposed for
a dynamical decaying $\Lambda$ with time \cite{f16}. Berman and collaborators \cite{f17} pointed out that
cosmological constant may vary as $ \Lambda \sim t^{-2}$,  which seems
to play a major role in cosmology. To obtain a time varying
cosmological constant,  one can assume different  phenomenological
relations proportional to  the Hubble parameter, one such  example \cite{f17}
considered in recent times is of the form  $\Lambda \propto H^2 $  where  $H$ is
Hubble parameter. Cosmological models with imperfect fluid have been explored  in the presence of 
 a dynamical cosmological constant and gravitational constant in Einstein gravity \cite{f18}. 
It is , therefore, important to look into viable cosmological  models  with  variable $G(t)$ and
    $\Lambda(t)$ in higher derivative theory of gravity in the presence of imperfect fluid. 

To describe a relativistic theory of viscosity, Eckart \cite{f19} made the first attempt. However, the theories of dissipation in 
 Eckart formulation  suffers from serious shortcoming, viz., causality and stability \cite{f20} regardless of the choice  of  equation 
of state. The problem arises due to first order nature of the theory, since it considers only first order deviation from equilibrium.
  It has been shown that the  problems of the relativistic imperfect fluid  may be resolved by including higher order deviation terms
 in the transport equation \cite{f21}. Israel and Stewart  \cite{f22}, and Pavon \cite{f23} developed a fully relativistic 
formulation of the theory taking into account second order deviation terms in the 
theory, which is termed as "transient" or "extended" irreversible thermodynamics (in short, 
{\it EIT}). The crucial difference between the standard Eckart and the extended Israel-Stewart transport equations is that the latter 
 is a differential evolution equations, while the former is an  algebraic relation. Extended irreversible thermodynamics takes its name
 from the fact that the set needed to describe non-equilibrium states is extended to include the dissipative variables ($\tau$, $\zeta$). 
     In irreversible thermodynamics, the entropy is no longer conserved, but grows, according to the second law of thermodynamics (i.e., $S^{\alpha}_{;\alpha}\geq 0)$. Bulk viscosity arises typically in mixtures either of different species or of the species but with different energies.
The solutions of the full causal theory are well behaved for all the times. Therefore, the best currently 
 available theory for analyzing dissipative processes in the universe
 is the full Israel-Stewart theory (FIS).  Using the transport equation obtained from {\it EIT}, in addition to dynamical equation obtained  from either  Einstein 
gravity \cite{f18, f25}  or modified gravity \cite{f24} cosmological solutions are generally obtained. 
 The motivation of this paper is to explore cosmological solutions  in the modified theory of gravity in the presence of  variable
$G(t)$ and $\Lambda(t)$  with imperfect  fluid described by FIS theory. It is interesting to study the behaviour of $G(t)$ and
 $\Lambda(t)$ in the presence of viscosity.\\
 The characteristic  
temperature of the  universe may   be determined in the presence of viscosity in the FIS theory. However, it may be pointed out here that
 in a number of literature \cite{f25}  the evolution of temperature of the universe in the presence of viscosity 
 from Gibbs equation is considered 
in the Einstein  gravity. We also consider the above procedure in the framework of higher order gravity to determined the behaviour of
 temperature. \\
The plan of this paper is as follows: in sec. 2, we give the
gravitational action and set up the relevant 
field equations in the higher derivative theory of gravity. In sec. 3,  
cosmological solutions are presented. In sec. 4,  
distance modulus curves  are presented. Finally, in sec. 5, we summarize the 
results obtained.%%%%%%%%%%%%%%%%%%%%%%%%%%%%%%%%%%%%%%%%%%%%%%%%%%%%% 
\section{\bf  Gravitational Action and Dynamical Equations:}
%%%%%%%%%%%%%%%%%%%%%%%%%%%%%%%%%%%%%%%%%%%%%%%%%%%%%
We consider a gravitational action with higher order term in the scalar curvature $(R)$ containing a variable gravitational constant $(G(t))$ which is given by 
\begin{equation}
{\large I} = - \int  \left[\frac{1}{16\pi G(t)} f(R)   + L_{m}  \right] \sqrt{- g} \; 
d^{4}x
\end{equation}
where $f(R)$ 
is a function of $ R$ and its higher power including a  variable cosmological 
constant $\Lambda(t)$, $g$ is the determinant of the four dimensional metric  and $ { L_m}$ represents the matter Lagrangian.\\
Variation of the action (1) with respect to  $g_{\mu\nu}$ yields
\[
f_R(R) \; R_{\mu\nu} - \frac{1}{2} \; f(R)\; g_{\mu\nu} + f_{RR} (R) \left( 
\nabla_{\mu} \nabla_{\nu} R -g_{\mu\nu}\nabla^\mu \nabla^\nu  g_{\mu\nu} \right) +
\]
\begin{equation}
f_{RRR} (R)\left( \nabla_{\mu}R \nabla_{\nu} R  - \nabla^{\sigma} R 
\nabla_{\sigma} R \; g_{\mu\nu}
\right) = - 8\pi G(t) \; T_{\mu\nu}
\end{equation}
where $ \nabla_\mu $ is  the
covariant differential operator,  $f_R (R)$ represents the derivative of $f(R)$ with 
respect to $ R$ and  $T_{\mu\nu}$ is the effective energy momentum tensor for matter 
determined by ${L_m}$.
We consider a flat Robertson-Walker spacetime given by the metric
\begin{equation}
ds^{2}=-dt^{2}+a^2(t)\left[dr^{2}+r^{2}(d{\theta}^{2}+sin^{2}{\theta } 
d{\phi}^{2})\right ]
\end{equation}
where $a(t)$ is the scale factor of the universe. The scalar curvature  for a flat universe is
\begin{equation}
R= - \; 6  \; [ \dot{H} + 2H^{2}]
\end{equation}
where  $ H=\frac{\dot{a}}{a} $ is the Hubble parameter and an  overdot represents 
derivative with respect to cosmic time $(t)$. The trace and (0,0) 
components of eq. (2) are given by
\begin{equation}
R f_R (R) - 2f(R) \;+\;3f_{RR}(R) \left(\ddot{R}+3 \frac{\dot{a}}{a} \dot{R} \right) + 
3 f_{RRR} (R) \dot{R} + 8 \pi G(t) \;T = 0,
\end{equation}
\begin{equation}
f_R(R) \; R_{00} +\frac{1}{2} f (R) -3f_{RR}(R) \frac{\dot{a}}{a}\dot{R} +8\pi G(t)\;  T_{00} =0 .
\end{equation}
Let us consider a  higher order gravity, namely,  $ f(R)= R +\alpha R^2 -2 \Lambda(t)$. Using Eq. (3) in Eqs. (5) and (6),  we get 
\begin{equation}
H^{2}-6{\alpha}\left[2H \ddot{H} -\dot{H}^{2} +6\dot{H} H^{2}\right]
= \frac{8 \pi G(t)\rho}{3}+\frac{\Lambda(t)}{3},
\end{equation}
 and the conservation equation becomes 
\begin{equation}
\dot{\rho}+ 3(\rho+p)H= - \left(\frac{\dot G}{G}\rho+\frac{\dot{\Lambda}}{8\pi G} \right), 
\end{equation}
where $\rho $ and $p$ are  the  energy density and pressure of the 
perfect fluid respectively. 
Equations    (7) and  (8) are the key equations to study cosmological
 models with a perfect fluid in the presence of time varying $G$ and $\Lambda$. To include,  the effect of viscosity   in the above, the perfect fluid pressure in eq. (8) is replaced by an effective pressure $ p_{\it {eff}} $, which is given by
$ p_{\it {eff}} = p + \Pi$, 
where  $p$ is isotropic pressure and $\Pi$ is the bulk viscous stress. In {\it $EIT$}, the bulk
viscous stress $\Pi$ satisfies a  transport equation  given by 
\begin{equation}
\Pi +\tau \dot{\Pi}=-3\zeta H -\frac{\epsilon}{2} \tau \Pi\left[3H + 
\frac{\dot\tau}{\tau}-\frac{\dot\zeta}{\zeta}-\frac{\dot T}{T}\right],
\end{equation}
where $\zeta$ is the coefficient of bulk viscosity, $\tau$ is the relaxation 
coefficient for transient bulk viscous effects and $T$ $\geq 0$ is the absolute temperature of the universe. The parameter $\epsilon$ takes the value 0 or 1. Here  $\epsilon = 0 $ represents  
truncated  Israel-Stewart  theory and $\epsilon = 1  $ represents
full Israel-Stewart  (FIS) causal 
theory. One recovers the non-causal Eckart theory for $\tau = 0$.   The conservation eq. (8) including viscous fluid is given by :
\begin{equation}
\dot{\rho}+ 3(\rho+p + \Pi)H= - \left(\frac{\dot G}{G}\rho+\frac{\dot{\Lambda}}{8\pi G} \right). 
\end{equation}
For a constant $G$, $\Lambda$  and $\Pi = 0$, eq. (10) reduces to
the usual  continuity equation for a barotropic fluid. 
We consider an  equation of state for the isotropic fluid pressure given by 
\begin{equation}
 p =(\gamma -1)\rho 
\end{equation}
where $\gamma $ $(1\leq\gamma\leq 2)$ is a constant. The deceleration parameter $(q)$ is related to $H$ as 
\begin{equation}
 q = \frac{d}{dt}\left(\frac{1}{H} \right) -1.
\end{equation}
The  deceleration parameter is negative  for accelerating and positive
for decelerating phase of the universe. The temperature of the universe  is defined via the Gibbs equation which is given by 
\begin{equation}
 TdS = d \left( \frac{\rho}{n} \right)+ p \; d \left({\frac{1}{n}}\right).
\end{equation}
The behaviour of temperature  in the universe  is obtained  through Gibbs integrability condition
\begin{equation}
 n\frac{\partial T}{\partial n} + \left(\rho + p \right)\frac{\partial T}{\partial \rho} = T 
 \frac{\partial p}{\partial \rho}.
\end{equation}
 For a barotropic fluid the temperature follows a  power-law which is  $ T \sim \rho^{\frac{\gamma -1}{\gamma}}$.  
 The above temperature may be  determined using  Gibbs integrability condition also 
 \begin{equation}
 \frac{\dot T}{T} = -3H \left[ \left(\frac{\partial p}{\partial \rho}\right)_n + \frac{\Pi}{T} \left(\frac{\partial T}{\partial \rho}\right)_n \right].
 \end{equation}
 To determine  the temperature of a viscous universe  eq. (9) may be employed here, in addition to, 
  Gibbs integrability condition (15). Both of them are considered here to  investigate viable cosmological models.
 %%%%%%%%%%%%%%%%%%%%%%%%%%%%%%%%%%%%%%%%%%%%%%%%%%%%%
\section{\bf Cosmological  Solutions :}
%%%%%%%%%%%%%%%%%%%%%%%%%%%%%%%%%%%%%%%%%%%%%%%%%%%%%
The system of eqs. (7), (9)-(11) is employed to obtain cosmological solutions. The system of equations
is not closed as it has eight unknowns ($\rho$, $\gamma$, $\tau$,  $\zeta$, $G$ 
$\Lambda$,  $a(t)$,  $T $) to be determined  from four  equations.
 We assume the following widely accepted $\it{ad hoc}$ relations
\begin{equation}
\zeta = \beta \rho^s, \; \tau = \beta \rho^{s-1}
\end{equation} 
where $\zeta \geq 0 $, $\tau \geq 0$ ,  $\beta \geq 0$ and $s\geq 0$. 
We consider in this paper a known  variation of $\Lambda $  with
Hubble parameter ($H$) of the form, $\Lambda = 3 m H^2 $ where 
$m$ is  dimensionless constant  to be determined from the dynamical
equations. In the next section  we explore cosmologies with
power law and exponential expansion respectively.
%%%%%%%%%%%%%%%%%%%%%%%%%%%%%%%%%%%%%%%%%%%%%%%%
{\subsection { \bf  Power-law  model:}}
%%%%%%%%%%%%%%%%%%%%%%%%%%%%%%%%%%%%%%%%%%%%%%%
In this case we consider a power law expansion of the universe given by 
\begin{equation}
a(t)= a_0 t^D
\end{equation}
where $a_0$ and $D$ are constants which are to be determined from the field equation.  The accelerating  mode of
 expansion ($q < 0$) of the universe 
 is obtained for $D> 1$. 
In the absence of particle creation ( i.e., $ n^{\alpha}_{;\alpha} =0$ ) eq. (10) may be decoupled as follows:
\begin{equation}
\dot{\rho}+\; 3\gamma \rho H +\;3\Pi H \;=  0,
\end{equation}
\begin{equation}
 8\pi \;\dot G\;\rho +\;\dot{\Lambda }\;= 0. 
\end{equation}
 Using  eqs. (7), (17) and (19) we obtain
\begin{equation}
G = G_0 \left[ t^2 + \rho_2 \right]^{\frac{m}{1-m}},  
\end{equation} 
 where we replace $\rho_2 = \frac{18\alpha (2D-1)}{1-m}$ and $ G_0 = const.$ The
 initial value of $G(t)$  is determined in terms of
 coupling parameter $(\alpha)$, which however vanishes when $D=\frac{1}{2}$.
 In this case the gravitational parameter $G(t)$ increases  with time 
for $\Lambda > 0$  and  it decreases for  $\Lambda < 0$.  It is
evident that the  model admits a constant 
   $G$  when  $\Lambda$ vanishes.
The  energy density evolves as
\begin{equation}
 \rho = \rho_0 \left[ t^2 + \rho_2 \right]^{\frac{1-2m}{1-m}} \; t^{-4},
\end{equation}
where  $\rho_0 = 3(1-m)D^2$ (with $8 \pi G_0 =1$). For physically realistic solution ( i.e., $\rho > 0$) the upper boundary on
 cosmological constant is $\Lambda < 3 H^2$ (i.e., $m <1 $).  For  $ m = \frac{1}{2} $  the variation of energy density
is determined by  $R^2$ term only for which it decreases as ($\rho
\sim t^{-4}$), whereas it is independent of $\alpha$. 
The   bulk viscous stress   obtained 
 from  eq. (18) is given by
\begin{equation}
\Pi = -   \left[ \Pi_0 t^2 + \Pi_2 \right] \left[ t^2 + \rho_2 \right]^{\frac{-m}{1-m}} t^{-4}  ,
\end{equation}
 where $\Pi_0 = D(3\gamma D (1-m)- 2)$ and $\Pi_2 = 18\alpha D(2D-1)(3\gamma D -4)$. 
 For physically  realistic solution bulk viscous stress is essentially
 negative,  which demands $\Pi_0> 0$ i.e., $D > \frac{2}{3\gamma
   (1-m)}$. We note that the bulk viscous
 stress decreases as $|\Pi|  \sim t^{- \frac{2}{1-m}}$
 when  (i) $\alpha =0$ and  $D \neq
 \frac{1}{2}$  or (ii) $\alpha \neq 0$ and  $D = \frac{1}{2}$. It is evident that for 
 $m <1$, bulk viscous stress ($|\Pi|$) will decrease  with time. One of the advantage of  the
 FIS theory is that one can determine the evolution of the temperature
 in this case.   For
 FIS theory we use $\epsilon =1$, eq. (9)  reduces to  a
 differential equation given by   
\begin{equation}
\frac{\dot{T}}{T} = 3H -\frac{\dot \rho}{\rho} + \frac{6H\rho}{\Pi}+ \rho^{1-s} \frac{2}{\beta} + \frac{2\dot{\Pi}}{\Pi}.
\end{equation}
Using  eqs. (17), (21) and (22) in eq. (23) we get temperature
 of the universe, which is 
\begin{equation}
T = T_0 \; \frac{\Pi^2 a^3}{\rho} e^{\int{\frac{6\rho H}{\Pi}}dt} e^{\frac{2}{\beta}\int{\rho^{1-s}}dt},
\end{equation}
where $T_0$ stands for a constant. The  temperature of the universe have to go through Gibbs integrability condition  (15) for a viable solution.
We note the following: \\
(i) For  $s=\frac{3}{4}$, $m= \frac{1}{2}$ and $\alpha = 0 $, we
note power law decrease of the temperature, which is  $T= T_0 \;t^{- \alpha_1 }$,
 where $\alpha_1 = \frac{6\rho_0 D}{\Pi_0}+ 4- 3D-\frac{2\rho_0^{\frac{1}{4}}}{\beta}$. The decreasing mode of temperature 
is ensured for $ \alpha_1  \geq  0$. Putting the expression of  temperature ($T$) in the presence of viscosity in Gibbs condition  (15) 
one can obtain  ${\alpha_1 =\frac{4(\gamma -1)}{\gamma}}$. Here the negative value of bulk viscous stress ($\Pi$) is obtain for 
$\Pi_0 > 0$ (i. e., $3\gamma D >4$).  However, In the absence of viscosity the variation of temperature can be obtained from eq.
 (15) which yields $T=T_0 t^{-3D(\gamma-1)}$. It is evident from fig. (1) that the temperature of the universe is more for a universe 
filled viscous fluid  at a given time compared to that of a universe without viscosity.  
\input{epsf}   
\begin{figure}[hbtp]
\begin{center}
\includegraphics[width=8cm]{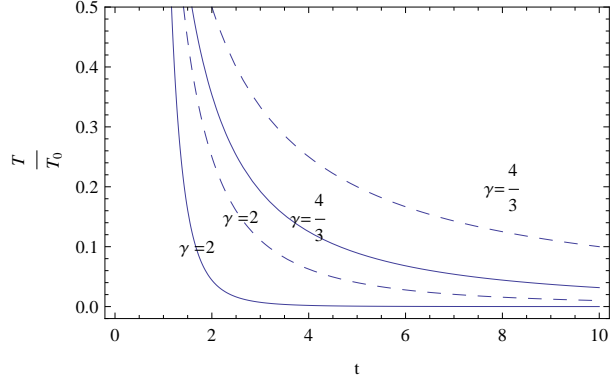}
    \caption[  shows the plot of  $\frac{T}{T_0}$  Vs.  $t$ for a different value of $\gamma$
 with $D=\frac{3}{2}$. Here solid line represent the variation in the absence  of viscosity while dashing line represent that in the presence of viscosity.]{\label{fig402}  shows the plot of  $\frac{T}{T_0}$  vs  $t$ for a different value of $\gamma$
 with $D=\frac{3}{2}$. Here the solid line represent the variation in the absence  of viscosity while the  dashing line represent that in the presence of viscosity.}
\end{center}
\end{figure} 
It is also evident that  the temperature of the universe is  higher for lower  values of $\gamma$ at a given instant of time.
   
(ii) For $s=\frac{3}{4}$, $ m=  \frac{5}{6}$, $\alpha \neq 0$ and
$D \neq \frac{4}{3 \gamma} $, the temperature 
  of the universe evolves as 
$T = T_0 t^{(3D - 4 - \frac{6\rho_0 \rho_2 D}{\Pi_2} + \frac{2\rho_0^{\frac{1}{4}}}{\beta \rho_2} )} 
 \left[\Pi_0 t^2 +\Pi_2 \right]^{(2-\frac{3\rho_0 D}{\Pi_0} + \frac{3\rho_0 \rho_2 D}{\Pi_2})}
 \left[t^2 +\rho_2\right]^{-6-\frac{\rho_0^{\frac{1}{4}}}{\beta
     \rho_2}}$.
  At the later stage of evolution of the universe ($ t>> \Pi_2$ and $t>> \rho_2$) one can obtain the power law variation of temperature. 
 Using   Gibbs integrability condition   the temperature evolution of the universe becomes $T= T_0 t^{\frac{-12(\gamma -1)}{\gamma}}$.
 Here the condition $ D>\frac{4}{\gamma}$ ensure the  negative value of bulk viscous stress.
   
(iii) For $s=\frac{m+1}{2}$,   the temperature of the universe 
 in GR ($\alpha = 0$) may be evaluated using Gibbs integrability condition
  (15) and eq. (9),  we obtain the following  variation of temperature : 
  \begin{equation}  
     T= T_0 t^{\frac{-2(\gamma -1 )}{\gamma(1-m)}},
   \end{equation} 
   which is decaying if $\gamma >1$, $m < 1$ or $\gamma < 1$, $m > 1$. Here the negative value of bulk viscous stress ($\Pi$) is obtain for $\Pi_0 > 0$
     ( i. e., $ D > \frac{4}{3\gamma(1-m)}$). The condition $ D > \frac{4}{3\gamma(1-m)}$ also  implies   that in an expanding universe the temperature decrease less rapidly in  the presence of viscosity  compared to that when viscosity is absent. The plot of $T$ Vs. $t$ in fig. (2) shows that at  a given time,  a  universe with higher temperature  is possible when the cosmological constant  is gradually decreased in GR. It permits a universe with late acceleration.\\  
\begin{figure}[hbtp]
\begin{center}
\includegraphics[width=8cm]{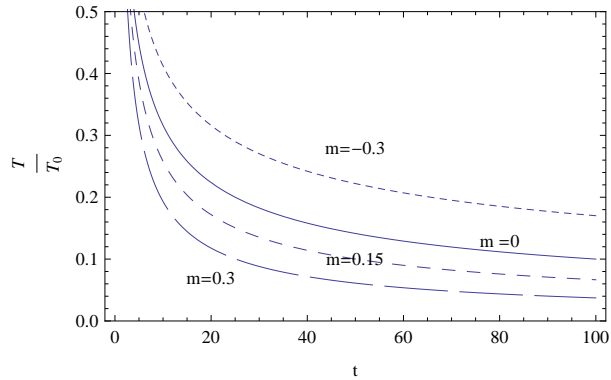}
    \caption[ shows the plot of  $\frac{T}{T_0}$  Vs.  $t$ for a different value of cosmological constant (i.e., m)  for $\gamma=\frac{4}{3}$.]{\label{fig403}  shows the plot of  $\frac{T}{T_0}$  Vs.  $t$ for a different value of  cosmological constant (i.e., m) for $\gamma=\frac{4}{3}$.}
\end{center}
\end{figure}
  Using the eq. (25),  One can obtain the temperature of the viscous universe at the different epoch of cosmological 
evolutions of the universe. To  obtain the temperature at present epoch we set the initial condition $T_0 = 1\times 10^{10} $ at $t=1$  \cite{f25a} i.e., during the cosmological epoch of decoupling of neutrinos from the cosmic plasma.
   Considering  age of the present universe $\sim 10^{10}$ years $\sim 3.15 \times 10^{17}$ second, $\gamma = \frac{4}{3}$ (radiation) and 
cosmological constant $\Lambda = 3\times0.085\times H^2 (i.e., m = 0.085)$, eq. (25) leads to  $T \sim 2.74$K which is in fair agreement 
with observed value $T \sim 2.72$K from CMBR. So, the solution admits observed  value of temperature from CMBR  in the presence 
of a small value of cosmological constant with an  late  accelerating universe.
 %%%%%%%%%%%%%%%%%%%%%%%%%%%%%%%%%%%%%%%%%%%%%%%% 
{\subsection{ \bf  Exponential models:}}
%%%%%%%%%%%%%%%%%%%%%%%%%%%%%%%%%%%%%%%%%%%%%%%
The set of  eqs. (7), (9), (18)-(19)  also admit  cosmological solution with a 
 universe without singularity . We discuss two such cases.\\
{\bf Case I} : We note that   Hubble
parameter  satisfying a differential equation
\begin{equation}
\dot{H} =  \eta H - \frac{3}{2} H^2,
\end{equation}
 is permitted with   $ \eta = \sqrt{\frac{1-m}{6\alpha }} $. This admits a ever expanding universe with no singularity, such a solution admits emergent universe scenario \cite{f27}. 
In this case the  corresponding variation of   gravitational
constant and energy density are given by  
\begin{equation}
 G = G_0 exp[b H^{-2}], 
\end{equation}
\begin{equation}
  \rho =\rho_0 H^4 exp[-b H^{-2}],
\end{equation}
where   $\rho_0 = \frac{81 \alpha}{2}$ (with $8\pi G_0 = 1 $) and $b=
\frac{2 m}{27\alpha}$ and $G_0 = const.$ It is evident that the model admits a constant $G$ when $\Lambda = 0 $. 
 The bulk viscous stress is obtained from eq. (18), which yields 
\begin{equation}
\Pi = (2-\gamma)\rho - \frac{1}{3} \rho H^{-3}\left[ 4  \eta H^2 + 2 b
  \eta  -3 b H \right].
\end{equation}
We note that without a cosmological constant ($ \Lambda = 0 $), one obtains 
a realistic solution when Hubble parameter satisfies an upper bound 
 $ H < \frac{4 \eta}{3(2-\gamma)}$.  
The temperature evolution  of the universe in full causal theory  is obtained from eq. (23), which yields
\begin{equation}
T = T_0 \;e^{f(H)}\; e^{g(H)} \;  \;
\frac{\Pi^2}{\rho ( H - \frac{ 2 \eta}{3})^{-2}    }
\end{equation}
where $f(H)=\int{\frac{6\rho H}{\Pi}} dt$ and $g(H)= \frac{2}{\beta}\int{\rho^{1-s}} dt$.
 The above expression of the temperature also have to satisfies the Gibbs integrability condition (15). 
 We note the following :\\
(i) For $ s=\frac{1}{2}$, $\beta =\frac{2 \sqrt{3}}{3}$, in the absence of
 cosmological constant and for stiff fluid the temperature of the universe 
 evolves as $T = T_0 H^2 (H-\frac{2\eta}{3})^{-2}$.
  When  $H \ll \frac{2 \eta}{3} $, the evolution of temperature  becomes  $T \sim H^2$. 
 We obtain here a decreasing mode of temperature  for an exponentially expanding universe.
  
The scale factor of the  universe is obtained on integrating eq. (26) which is given by 
\begin{equation}
a(t)_{\pm}=\left[ a_1  \pm  a_{2}\; e^{\eta t} \right]^{\frac{2}{3}}.
\end{equation}
 The  solution ($a_+$) is important for
building emergent universe scenario \cite{f27}. It has no 
singularity  and the universe originated from a static state in 
the infinite asymptotic past ($t \rightarrow - \infty$). In this case, we note that the solution ($a(t)= a(t)_+$) represents a 
 universe which begins with a finite size in the past and grows
 exponentially. However, initially at  $t \sim 0$,  the universe is matter
 dominated which subsequently emerges to an  accelerated phase of expansion. 
 We note that the temperature of the universe decreases in this case. The
 decreasing mode of temperature obtained  in the FIS theory  may be
 relevant for the later evolution which will be taken up else where for a detail scenario. 
  \\
 {\bf Case  II :} The dynamical equation admit  de Sitter solution with
 $H =  \frac{ 2 \eta }{3}$.  The scale factor of the universe evolves as $a(t) =
a_0 e^{\frac{2 \eta}{3} t} $, for a sufficient inflation to solve cosmological
problems, de Sitter phase should exit after an epoch  $ \Delta t > \frac{195}{2
  \eta}$. We note de Sitter phase with 
\begin{equation}
 G=const.,\; \rho = const.,\; \Lambda = const.
\end{equation}
 and $\zeta = const.$ and $\tau = const.$ for  $m \neq 1$.
  In this case that bulk viscosity also remains constant throughout  the  inflationary phase. 
The temperature of the universe in this case is found to be a  constant ( similar to that obtained by A. Beesham in {\it Ref}  \cite{f18}) 
 for  $\beta = \frac{\gamma \;\rho^{1-s}}{(2-\gamma) \eta}$.
 
We also note that  for $ H = \frac{2}{3} \eta $ and $m = 1$, one ends up with
$\Lambda = const., $ $G= const.,$ $\rho =0,$ $ \zeta =0$, $ \tau = 0$
and $T=0$, which is not physically relevant.
%%%%%%%%%%%%%%%%%%%%%%%%%%%%%%%%%%%%%%%%%%%%%%%%%%%%%%
\section{\bf  Distance   Modulus  Curves :}
%%%%%%%%%%%%%%%%%%%%%%%%%%%%%%%%%%%%%%%%%%%%%%%%%%%%%%%%
We now probe late universe with  exponential  and power law expansion  taking into account observational 
results.  The distance modulus is $\mu = 5\log d_L +25$, where the luminosity distance $d_L  = r_1(1+z)\; a(t_0)$ and $z$ represents the red shift parameter, where   $1+z = \frac{a(t_0)}{a(t_1)}$. We determine 
  $r_1$  from  
\begin{equation}
  \int\limits_{0}^{r_1}  \frac{dr}{\sqrt{1-k r^2}} = \int\limits_{t_1}^{t_0} \frac{dt}{a(t)} .
\end{equation} 
At the late time,  the scale factor  becomes  $a(t) \sim a_2 e^{\frac{2\eta t }{3}}$.  For exponential expansion of the  flat universe, the distance modulus relation is given by  
\begin{equation}
\mu(z) = 5\log(\frac{z(1+z)}{H_0}) +25 ,
\end{equation}
 where $H_0 =\frac{2\eta}{3}$. 
For power law expansion $(a(t)\sim t^D)$ with $k=0$,  the distance modulus relation is given by 
\begin{equation}
\mu(z) = 5\log \left((\frac{D}{H_0})^\frac{1}{D}  \frac{(1+z)}{D-1} ((1+z)^\frac{D-1}{D}-1)\right) +25.
\end{equation}
The observed  values of $\mu(z)$ at different $z$ parameters \cite{f26} given in table 1 are employed to draw the curves corresponding to the
  exponential and power law expansion of the universe discussed above. The plot are shown in figs. (3) and fig. (4) which  matches  with observations perfectly.  
 \begin{figure}[hbtp]
\begin{center}
\includegraphics[,width=8cm]{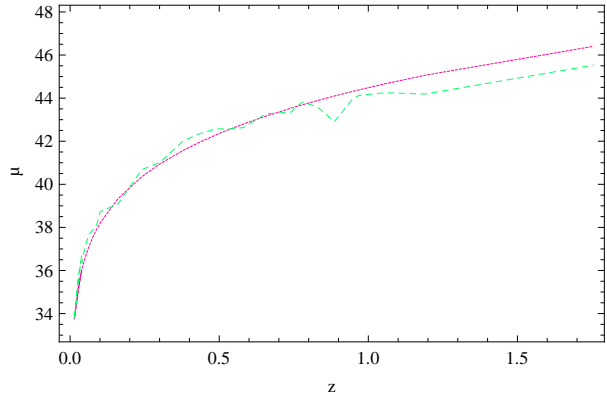}
    \caption[ shows the plot of  $\mu$  Vs.  $z$ for supernova data (dashing line) and for the exponential expansion  (solid line) with $ \eta = 0.00038$.]{\label{fig46}  shows the plot of  $\mu$  Vs.  $z$ for supernova data (dashing line) and for the exponential expansion  (solid line) with $ \eta = 0.00038$.}
\end{center}
\end{figure}
\begin{figure}[hbtp]
\begin{center}
\includegraphics[,width=8cm]{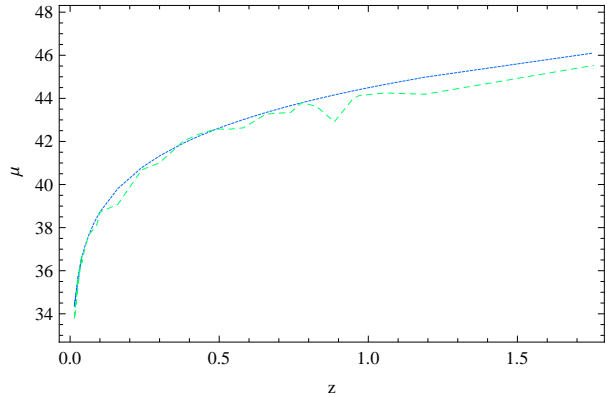}
    \caption[  shows the plot of   $\mu$  Vs.  $z$ for supernova data (dashing line) and for the power law expansion (solid line)  with $ H_0 =10^{-5}$ and $D= \frac{4}{3}$.]{\label{fig47} shows the plot of   $\mu$  Vs.  $z$ for supernova data (dashing line) and for the power law expansion (solid line)  with $ H_0 =10^{-5}$ and $D= \frac{4}{3}$.}
\end{center}
\end{figure}
\begin{table}[htbp]
 \begin{center}
 \begin{tabular}{|l|c|c|r|} \hline
z & supernova $\mu$&Exponential $\mu$ & Power law $\mu$\\ \hline
0.038 & 36.67 & 35.961 &36.543 \\ \hline 
0.014 & 33.73 & 33.742 &34.343 \\ \hline 
0.026 & 35.62 & 35.112 &35.703 \\ \hline 
0.036 & 36.39 & 35.839 &36.423 \\ \hline 
0.040 & 36.38 & 36.077 &36.657 \\ \hline 
0.050 & 37.08 & 36.582 &37.154 \\ \hline 
0.063 & 37.67 & 37.110 &37.673 \\ \hline 
0.079 & 37.94 & 37.634 &38.184 \\ \hline 
0.088 & 38.07 & 37.887 &38.43 \\ \hline 
0.101 & 38.73 & 38.212 &38.745 \\ \hline 
0.160 & 39.08 & 39.324 &39.814 \\ \hline 
0.240 & 40.68 & 40.349 &40.783 \\ \hline 
0.300 & 41.01 & 40.936 &41.329 \\ \hline 
0.380 & 42.02 & 41.579 &41.921 \\ \hline 
0.430 & 42.33 & 41.925 &42.235 \\ \hline 
0.490 & 42.58 & 42.298 &42.572 \\ \hline 
0.526 & 42.56 & 42.504 &42.757 \\ \hline 
0.581 & 42.63 & 42.797 &43.018 \\ \hline 
0.657 & 43.27 & 43.165 &43.345 \\ \hline 
0.740 & 43.35 & 43.530 &43.665 \\ \hline 
0.778 & 43.81 & 43.686 &43.801 \\ \hline 
0.828 & 43.61 & 43.881 &43.971 \\ \hline 
0.886 & 42.91 & 44.096 &44.158 \\ \hline 
0.949 & 43.99 & 44.316 &44.348 \\ \hline 
0.970 & 44.13 & 44.388 &44.490 \\ \hline 
1.056 & 44.25 & 44.665 &44.646 \\ \hline 
1.190 & 44.19 & 45.061 &44.983 \\ \hline 
1.755 & 45.53 &46.403  &46.104 \\ \hline 
 \end{tabular}
    \end{center}
    \caption[Red Shift data from Supernova]{\label{Red shift data from  Supernova explosion} Red shift data from Supernova,  power law model and exponential model.}
\end{table}

 \pagebreak
 %%%%%%%%%%%%%%%%%%%%%%%%%%%%%%%%
\section{\bf   Discussions :}
%%%%%%%%%%%%%%%%%%%%%%%%%%%%%%%%
In this paper we study  both power law and exponential behaviour of the universe with viscosity 
 separately in a higher
derivative theory of gravity considering a time varying cosmological and gravitational constant. 
 We note that for a power law evolution of the universe, one obtains an increasing mode of  gravitational constant with a 
 positive  $\Lambda$ but  a decreasing mode of gravitational constant results for  a  
 negative $\Lambda$. We note a  physically realistic ($\rho_0 > 0$) solution for $\Lambda < 3 H^2$.
 For $\Lambda = \frac{3}{2} H^2 $  the variation of energy density ($\rho \sim t^{-4}$) is determined by $R^2$ term only. However,
 the energy density does not depend on the coupling parameter $\alpha$ in the gravitational action. We determine the characteristics 
 temperature of a viscous universe from  eq. (9), which also follows 
 from Gibbs integrability condition. Fig. (1) shows the variation of temperature for different values of $\gamma$  in the presence or 
in the  absence of viscosity in an  accelerating universe $(q>0)$. It is evident that the higher value of $\gamma$ leads to a universe
 with lower temperature at a given instant of time. The evolution of  temperature of a viscous universe is found to be more than that 
in a universe without viscosity.   Fig. (2) shows the variation of temperature for different values of $\Lambda$ (determined by $m$). 
It is evident from fig. (2) that  the rate of decrease of temperature is higher for larger values of cosmological constant in the presence
  of viscosity. Here  we obtain a interesting solution which suggests that the  present temperature of the universe is  $T\sim 2.745$K,
 which is in fair agreement with observed value $T \sim 2.72$K from CMBR, for a radiation dominated late accelerating  universe 
in the presence of a positive  cosmological ($\Lambda = 3\times0.085\times H^2$)  in GR. 
 We also note   cosmological solutions which admit a universe originating from
singularity free state. One of the  solution corresponds to emergent
universe \cite{f27} which  is interesting. We obtain a de Sitter
solution with $\Lambda \neq 3 H^2$. The  figs. (3)-(4) show the  plot of  distance modulus ($\mu$) vs red  shift parameter ($z$). The figures  indicate that  power-law and exponential model support the present  observational data perfectly well.
 
\end{document}